\documentclass[aps,prl,twocolumn,superscriptaddress,showpacs,preprintnumbers,amsmath,amssymb]{revtex4}
\usepackage{graphicx}
\usepackage{dcolumn}
\usepackage{bm}

\newcommand{\moeller}{M\o ller}
\newcommand{\cherenkov}{Cerenkov}
\newcommand{\solidangle}{0.62}      
\newcommand{\beamenergyhighII}{855.15} 
\newcommand{\beamenergylow}{569.31}   

\newcommand{\transmissioncomptonangleaccuracyhigh}{1.6}
\newcommand{\transmissioncomptonangleaccuracylow}{0.9}

\newcommand{\numberofchannelsforhighbeamenergy}{756}
\newcommand{\qsquaredaveragedhighI}{0.230}       
\newcommand{\qsquaredaveragedhighII}{0.230}       
\newcommand{\qsquaredaveragedlow}{0.106}       

\newcommand{\datahourslow}{54}
\newcommand{\datahourshigh}{46}

\newcommand{\experimentalasymmetryhigherrorstat}{2.31}

\newcommand{\experimentalasymmetrylowerrorstat}{0.89}

\newcommand{\experimentalasymmetryaluhighcorr}{-8.52}
\newcommand{\experimentalasymmetryalulowcorr}{-8.59}

\newcommand{\combinedsyspolerrorhighalucor}{0.87}

\newcommand{\combinedsyspolerrorlowalucor}{0.75}

\newcommand{\statisticalerrorlow}{\experimentalasymmetrylowerrorstat}               
\newcommand{\statisticalerrorhigh}{\experimentalasymmetryhigherrorstat}               

\begin{document}
\title{Measurement of the Transverse Beam Spin Asymmetry in Elastic
Electron Proton Scattering and the Inelastic Contribution to
the Imaginary Part of the Two-Photon Exchange Amplitude}
\author{F.~E.~Maas}
\email[corresponding author: ]{ maas@kph.uni-mainz.de}
\author{K.~Aulenbacher}
\author{S.~Baunack}
\author{L.~Capozza}
\author{J.~Diefenbach}
\author{B.~Gl{\"a}ser}
\author{Y.~Imai}
\author{T.~Hammel}
\author{D.~von Harrach}
\author{E.-M.~Kabu{\ss}}
\author{R.~Kothe}
\author{J.~H.~Lee}
\author{A.~Sanchez-Lorente}
\author{E.~Schilling}
\author{D.~Schwaab}
\author{G.~Stephan}
\author{G.~Weber}
\author{C.~Weinrich}
\affiliation{Institut f\"ur Kernphysik,
             Johannes Gutenberg-Universit{\"a}t Mainz,
             J.-J.-Becherweg 45,
             D-55099 Mainz,
             Germany}
\author{I.~Altarev}
\affiliation{St.~Petersburg Institute of Nuclear Physics, Gatchina, Russia}
\author{J.~Arvieux}
\author{M.~Elyakoubi}
\author{R.~Frascaria}
\author{R.~Kunne}
\author{M.~Morlet}
\author{S.~Ong}
\author{J.~Vandewiele}
\affiliation{Institut de Physique Nucleaire,
             91406 - Orsay Cedex,
             France}
\author{S.~Kowalski}
\author{R.~Suleiman}
\author{S.~Taylor}
\affiliation{
             Laboratory for Nuclear Science,
             Massachusetts Institute of Technology,
             Cambridge, MA 02139, USA}
\date{\today}
\begin{abstract}
We report on a measurement of the asymmetry in the scattering
of transversely polarized electrons off unpolarized protons, A$_\perp$, at two Q$^2$
values of \qsquaredaveragedlow~(GeV/c)$^2$ and \qsquaredaveragedhighII~(GeV/c)$^2$
and a scattering angle of $30^\circ < \theta_e < 40^\circ$.
The measured transverse asymmetries are
A$_{\perp}$(Q$^2$~=~\qsquaredaveragedlow~(GeV/c)$^2$)~=~(\experimentalasymmetryalulowcorr~$\pm$~\statisticalerrorlow$_{\rm stat}$~$\pm$~\combinedsyspolerrorlowalucor$_{\rm sys}$)~$\times$~10$^{-6}$
and
A$_{\perp}$(Q$^2$~=~\qsquaredaveragedhighII~(GeV/c)$^2$)~=~(\experimentalasymmetryaluhighcorr~$\pm$~\statisticalerrorhigh$_{\rm stat}$~$\pm$~\combinedsyspolerrorhighalucor$_{\rm sys}$)~$\times$~10$^{-6}$.
The first errors
denotes the statistical error and the second the systematic uncertainties.
A$_\perp$ arises from the imaginary part of the two-photon exchange amplitude
and is zero in the one-photon exchange approximation. From comparison with
theoretical estimates of A$_\perp$ we conclude that $\pi$N-intermediate states
give a substantial contribution to the imaginary part of the
two-photon amplitude. The contribution from the ground state proton
to the imaginary part of the two-photon exchange can be neglected.
There is no obvious reason why this should be different for the real part of the two-photon
amplitude, which enters into the radiative corrections for the Rosenbluth separation
measurements of the electric form factor of the proton.
\end{abstract}
%
\maketitle
\begin{figure}[htb]
\centering
  \includegraphics[width=0.2\textwidth]{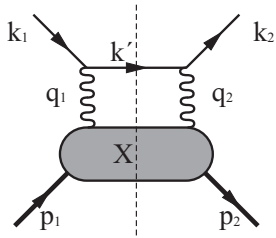}\\
  \caption{The two-photon exchange diagram. The filled blob X represents
           the response of the nucleon to the scattering of the virtual photon.}\label{fig:twophoton:feynman}
\end{figure}

%
%
The simple interpretation of electromagnetic probe experiments like elastic scattering of
electrons off protons is due to the smallness of the electromagnetic coupling constant
$\alpha \approx 1/137$ which allows to approximate
the electromagnetic transition amplitude as a single photon exchange process (Born approximation).
Higher order processes are treated as small \lq\lq radiative corrections\rq\rq\
like the two-photon exchange which is schematically shown in Fig.~\ref{fig:twophoton:feynman}.
It involves the exchange of two virtual photons (bosons) and an intermediate hadronic
state which includes the ground-state and all excited states of the hadronic system, which can be
off-shell for the real part of this box diagram amplitude. This makes the theoretical computation of the
two-photon effects difficult.
%
%
Tests of the limits of the validity of the one-photon approximation
have been done in the past, using different methods, like comparison of the $e^+p$ \
and $e^-p$ \ cross section data,
$\epsilon$-linearity of the ratio $R^2= (\mu_p G_E^p/G_M^p)^2$ in the Rosenbluth formula or observation
of T-odd polarization observables \cite{radcorQED:DeRujula:1971}. No effect has been found within the
accuracy of the experiments.
This discussion has been reactivated recently by the observation that the ratio of the
electric form factor of the proton to the magnetic form factor, $R=(\mu_p G_E^p/G_M^p)$,
is different when measured by the method of Rosenbluth separation as compared to the extraction
from polarization transfer.
The determination of the ratio $R$ from longitudinal-transverse (LT) or Rosenbluth separation
yields a value for $R$ which is consistent with $R \approx 1$~\cite{rosenbluth:arnold:1975,
rosenbluth:walker:1994,rosenbluth:andivahis:1994,rosenbluth:arrington:2003} in a Q$^2$ range $< 6$~(GeV/c)$^2$.
Recent polarization transfer measurements at Jefferson Laboratory \cite{polarisationtransfer:jones:2000,
polarisationtransfer:Gayou:2002} measure R from the ratio of the transverse to longitudinal polarizations
of the recoil proton, yielding a very different result $R \approx 1 - 0.135 (x - 0.24)$ where
$ x = $Q$^2$ in units of (GeV/c)$^2$.
It has been suggested
\cite{2photonexplanation:Guichon:2003,2photonexplanation:Brodsky:2003}
that a contribution from two-photon exchange can explain such a discrepancy.
There are observables which are directly sensitive to two-photon effects, like the transverse asymmetry
$A_\perp$ in the elastic scattering of transversely polarized electrons off unpolarized nucleons.
$A_\perp$ arises from the interference of the one-photon with the
two-photon exchange amplitude and is zero in Born approximation.\\
The treatment of the exchange of many photons is done in a framework similar to
the one developed for elastic $np$-scattering \cite{npscattering:Goldberger:1957}.
The parametrization of the scattering-amplitude consists of a set of six complex
functions, e.g. $\hat{\mathrm{G}}_{M}(s, Q^2)$,$\hat{\mathrm{G}}_{E}(s, Q^2)$, and
$\hat{\mathrm{F}}_{i}(s, Q^2),i=3 ... 6$,
which are generalized form factors.
The evaluation of the elastic cross section $d\sigma/d\Omega$ for the scattering of
electrons off protons has been discussed as well as quantities
like the polarization transfer from the electron to the nucleon, $P_l$ and $P_t$,
the electron-positron beam charge asymmetry, the target recoil normal spin
asymmetry, the transverse beam spin asymmetry ($A_\perp$), the depolarization tensor
and other variables
\cite{2photonamplitude:vanderhaeghen:2003,2photonamplitude:rekalo:2003-1,2photonamplitude:rekalo:2004,2photonaxial:drell:1965,2photonamplitude:rekalo:2003-2}.
For example the differential cross section for elastic electron-nucleon scattering
can be expressed as:
\begin{eqnarray}
    \frac{d\sigma}{d\Omega} &=& \sigma_0 \: \: \{ |\hat{\mathrm{G}}_{M}|^2 + \frac{\epsilon}{\tau}|\hat{\mathrm{G}}_{E}|^2
                      + 2 \epsilon \sqrt{\tau(1+\tau)\frac{1+\epsilon}{1-\epsilon}} \nonumber \\
                            & &[\frac{1}{\tau} |\hat{\mathrm{G}}_E| + |\hat{\mathrm{G}}_M|]{\cal R}(\hat{\mathrm{F}}_{3}(s, Q^2)) + {\cal O}(e^4)\}
                            \label{eq:crosssection}
\end{eqnarray}
The two-photon contribution appears in the real part of the amplitude
${\cal R}(\hat{\mathrm{F}}_{3}(s, Q^2))$.
An ab initio calculation of the real part of $\hat{\mathrm{F}}_{3}(s, Q^2)$
is at present impossible. It would require the knowledge of the off-shell form factors of the proton
in the intermediate state and all possible excitation amplitudes
for the intermediate state and their off-shell transition form factors.
A recent model calculation gives a contribution to the cross section
on the order of a few percent \cite{2photon:blunden:2003}. The authors
used the ad hoc assumptions that the intermediate state is described by an
on-shell particle and by the ground state only. A
parton model calculation which is applicable at the high Q$^2$ employed for the Rosenbluth
data \cite{2photonexplanation:vanderhaeghen:2004} yields a quantitative agreement with the
polarization transfer measurements.\\
As only the imaginary part of the two-photon amplitude
contributes via the interference with the one-photon
exchange amplitude \cite{radcorQED:DeRujula:1971} to $A_\perp$,
$A_\perp$ is proportional to the imaginary part of the combination of
$\hat{\mathrm{F}}_{3}(s, Q^2)$, $\hat{\mathrm{F}}_{4}(s, Q^2)$, and $\hat{\mathrm{F}}_{5}(s, Q^2)$.
The evaluation of $A_\perp$ yields \cite{2photon:gorchtein:2004, 2photonamplitude:vanderhaeghen:2003}:
\begin{eqnarray}
   A_\perp =  \frac{m_e}{M}\sqrt{2 \epsilon (1 - \epsilon)} \frac{\sqrt{1 + \tau}}{\tau}
   (1 + \frac{\epsilon}{\tau}\frac{\mathrm{G}_{E}^2}{\mathrm{G}_{M}^2})^{-1} \nonumber \\
  \times \: ( - \tau \: {\cal I}(\frac{\hat{\mathrm{F}_{3}}}{\mathrm{G}_M})
     - \frac{\mathrm{G}_{E}}{\mathrm{G}_{M}}{\cal I}(\frac{\hat{\mathrm{F}}_{4}}{\mathrm{G}_M}) \nonumber \\
    - \frac{1}{1+\tau}(\tau + \frac{\mathrm{G}_{E}}{\mathrm{G}_{M}}){\cal I}(\frac{\nu \hat{\mathrm{F}}_{5}}{M^2 \mathrm{G}_M})
  ) + {\cal O}(e^4).  \label{eq:aperp}
\end{eqnarray}
${\cal I}(\hat{\mathrm{F}}_{i}(s, Q^2))$ denotes the imaginary part of
$\hat{\mathrm{F}}_{i}(s, Q^2)$ and $\nu$ is the energy transfer to the proton. The order
of magnitude of $A_\perp$ is given by the factor $m_e/M \approx 5 \times 10^{-4}$.
At present, there is little information from experiments concerning
$\hat{\mathrm{F}}_{3}(s, Q^2)$, $\hat{\mathrm{F}}_{4}(s, Q^2)$, and
$\hat{\mathrm{F}}_{5}(s, Q^2)$.\\
In contrast to the real part of the two-photon exchange contribution, the
imaginary part of the two-photon amplitude can be calculated from the
absorptive part of the doubly virtual Compton scattering tensor with
two space-like photons \cite{2photonamplitude:vanderhaeghen:2003}.
The momenta of the boson and fermion in the loop
are given by momentum conservation. All intermediate hadronic states,
which can be excited due to the kinematics, contribute to $A_\perp$.
The calculation of $A_\perp$ on the proton at low Q$^2$ requires known
quantities, like elastic scattering form factors of the proton (elastic contribution) and
transition amplitudes to $\pi$N-intermediate states (inelastic contribution).
The SAMPLE collaboration has recently reported on the first measurement of
A$_\perp$ at a laboratory scattering angle of $130^\circ < \theta_e < 170^\circ$
and a Q$^2$ of 0.1~(Gev/c)$^2$ \cite{sample:wells:00}.
%
%
%
We report here on a measurement of $A_\perp$ at
similar Q$^2$, but much higher energy, and at forward angle.
Thus, we are not only sensitive to
the ground state as in the case of the SAMPLE measurements, but also to
$\pi$N-intermediate states. In addition, both photons are
space like in forward scattering while in contrast at backward angles
the asymmetry is dominated by cases where one of the photons is
quasi real \cite{2photonamplitude:vanderhaeghen:2003}.
$A_\perp$ is an asymmetry in the cross section
for the elastic scattering of electrons with spin parallel ($\sigma_\uparrow$)
and spin anti-parallel ($\sigma_\downarrow$) to the normal scattering vector
defined by $\vec{S}_n = (\vec{k}_e \times \vec{k}_out)/|\vec{k}_e \times \vec{k}_{out}|$.
$\vec{k}_e$ and $\vec{k}_{out}$ are the three-momentum  vectors of the initial and final
electron state. The measured asymmetry $A_\perp^m$ can be written as
$A_\perp^m$~=~($\sigma_\uparrow - \sigma_\downarrow)/(\sigma_\uparrow + \sigma_\downarrow)$~=~$A_\perp \vec{P}_e \cdot \vec{S}_n$.
$A_\perp$ is a function of the scattering angle $\theta_e$, the four-momentum transfer
Q$^2$ and the electron beam energy E$_e$. The term $\vec{P}_e \cdot \vec{S}_n$ introduces a
dependence of $A_\perp^m$ on the electron azimuthal scattering angle $\phi_e$ with a zero
crossing for the case where the scattering plane contains the incident electron polarization vector $\vec{P}_e$.
$A_\perp^m$ vanishes for $\theta_e = 0^\circ$ (forward scattering)
and for $\theta_e = 180^\circ$ (backward scattering). It vanishes also if the electron
polarization vector is longitudinal.
\begin{figure}
  \centering
  \includegraphics[width=0.15\textwidth]{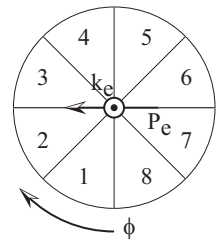}\\
  \caption{The momentum vector $\vec{k}_{e}$ is pointing here out of the paper plane.
           The momentum vector $\vec{k}_{out}$ of the outgoing electron
           can take all possible $\phi_e$ values. Both together
           define the coordinate system  according to the Madison convention
           \cite{madisonconvention:1970} with $\vec{S}_n = (\vec{k}_e \times \vec{k}_{out})/|\vec{k}_e \times \vec{k}_{out}|$.
           The direction of the electron polarization vector $\vec{P}_e$ for the + helicity state
           is indicated by the arrow.
           $\phi_e$ and $\phi_{\vec{P}_e}$ are counted as indicated.
           The elastic scattered electrons are detected in the $\phi_e$-symmetric
           PbF$_2$-calorimeter of the A4 experiment. For the extraction of
           $A_\perp^m$, the detector has been divided into 8 sectors
           as indicated in the figure.
           }\label{fig:aperp:angle}
\end{figure}
Fig.~\ref{fig:aperp:angle} shows a schematic defining $\phi_e$, $\vec{S}_n$,
and other quantities.\\
%
We have used the apparatus of the A4 experiment at the MAMI accelerator in Mainz
\begin{figure}
  \centering
  \includegraphics[width=0.35\textwidth]{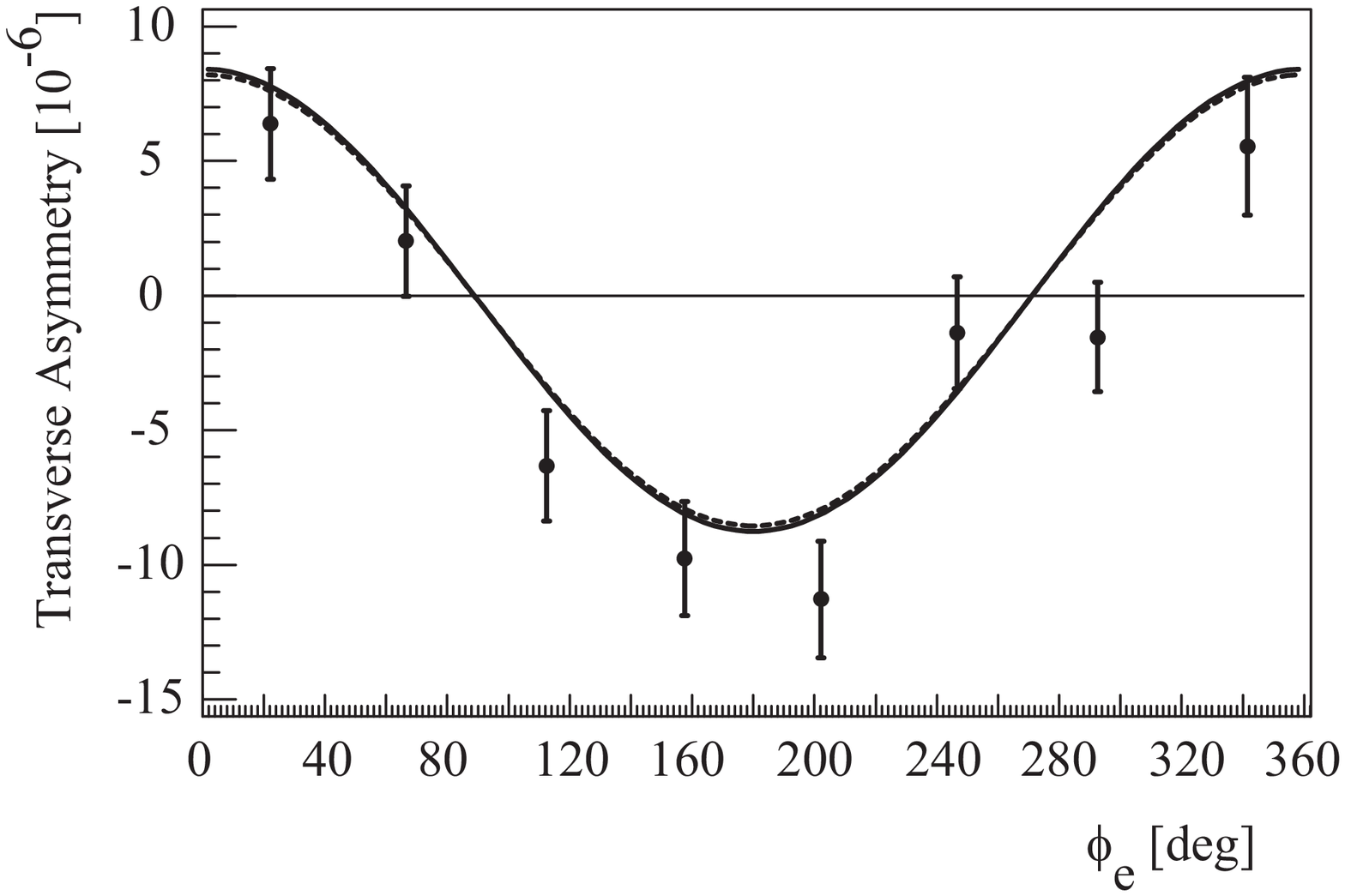}\\
  \includegraphics[width=0.35\textwidth]{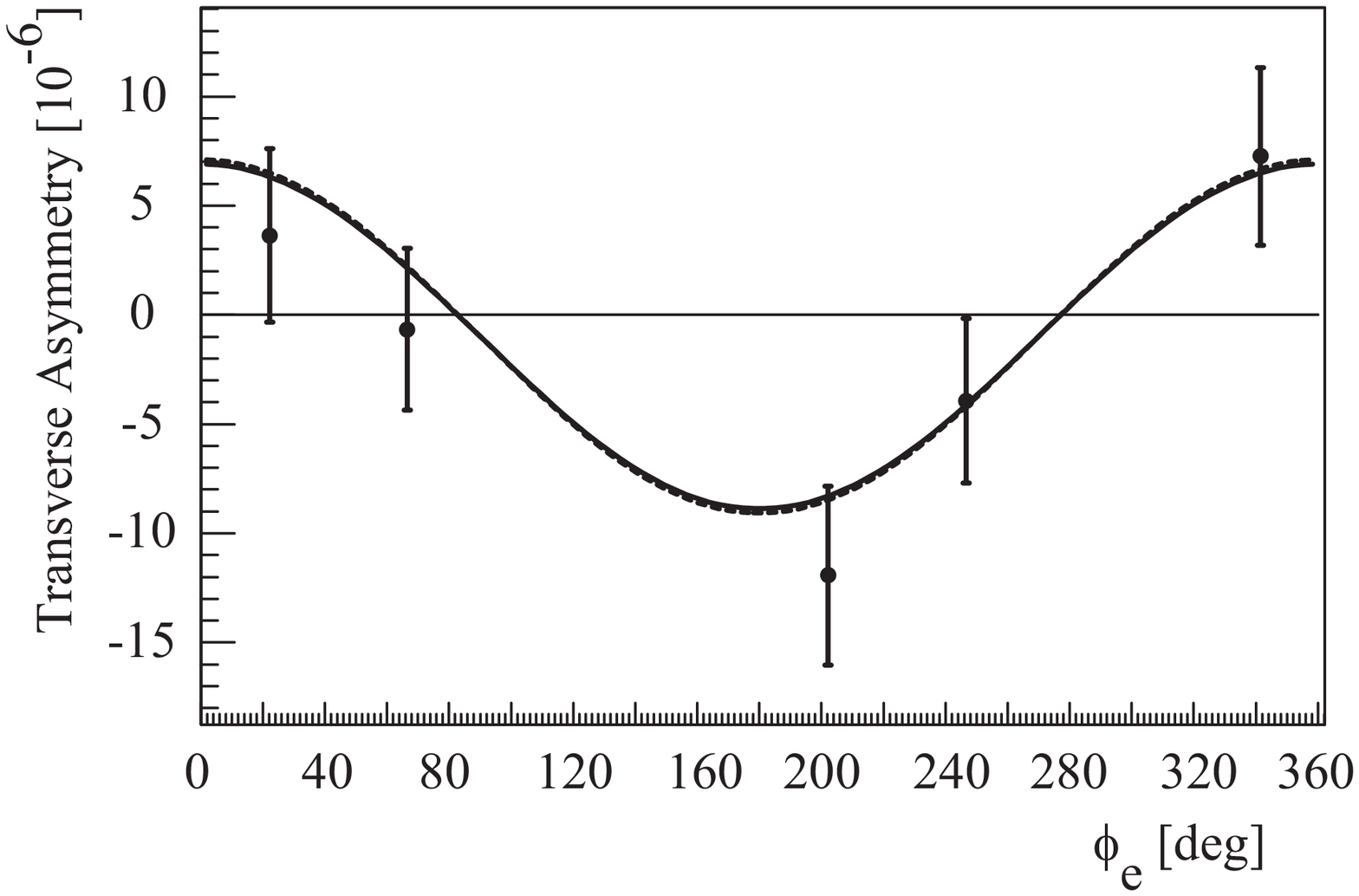}\\
  \caption{The upper plot shows the result of our measurements of $A_\perp$
           for a beam energy of \beamenergylow~MeV and at the lower plot for the beam energy
           of \beamenergyhighII~MeV. The asymmetry is plotted as a function of the laboratory angle
           $\phi_e$ as defined in Fig.~\ref{fig:aperp:angle}. The 1022
           PbF$_2$-detectors of the calorimeter have been divided into 8
           subsets (sector 1 to 8), each spanning an angular range of $\approx$45$^\circ$ in $\phi_e$.
           For the lower plot only \numberofchannelsforhighbeamenergy\
           channels of the PbF$_2$-detector had been installed (sector 1, 2, 5, 6, and part of sector 8).
           }\label{fig:aperp:measurement}
\end{figure}
to make a measurement of the transverse beam spin asymmetry
$A_\perp$ \cite{MAMI:euteneuer:94,a4calori:maas:02,a4experi:maas:03,a4experiment:maas:2004}.
%
%
The polarized electrons were produced using a strained layer
GaAs crystal which is illuminated with circularly polarized laser
light \cite{polsource:aulenbacher:97} resulting in longitudinally polarized
electrons. The sign of the electron beam polarization was switched
between the two patterns $(+--+)$ and $(-++-)$ randomly by means of a fast
Pockels cell in the optical system of the polarized electron source.
Average beam polarization was about $80\,\%$ which has been measured using a
\moeller\ polarimeter in a different experimental hall. The longitudinal spin
of the electrons leaving the photocathode has been rotated in the accelerator
plane using a Wien filter located between the 100~keV polarized electron source
and the injector linac of the accelerator. In addition the energy of the accelerator
has been tuned so that the relativistic spin precession in the three microtron
stages of the accelerator in combination with the Wien filter resulted in a
beam polarization perpendicular to the beam direction.
The rotation of the spin angle at the location of the experiment has been measured
using the transmission Compton polarimeter located between the liquid hydrogen
target and the electron beam dump.\\
The measurements of $A_\perp^m$ have been done with the Wien filter set
so that the electron polarization vector $\vec{P}_e$ shows for the $+$ helicity
to the negative
x-axis of a right handed coordinate system according to the Madison convention
\cite{madisonconvention:1970} and as indicated in Fig.~\ref{fig:aperp:angle},
corresponding to $\phi_{\vec{P}_e}=90^\circ$ and $\theta_{\vec{P}_e}=90^\circ$.
In this case the sign of $A_\perp^m$ as measured in sectors 4 and 5 (corresponding
to $\phi_{\vec{P}_e}=180^\circ$) is the same as $A_\perp$ and the sign of $A_\perp^m$
as measured in sectors 1 and 8 is opposite to $A_\perp$.
The transmission Compton polarimeter allowed
to set the angle of the beam polarization vector to an accuracy of
$\delta \theta_{\vec{P}_e}= \pm \transmissioncomptonangleaccuracyhigh^\circ$
and $\delta \theta_{\vec{P}_e}= \pm \transmissioncomptonangleaccuracylow^\circ$
for the beam energy of \beamenergyhighII~MeV and \beamenergylow~MeV, respectively.
%
%
For the measurements of $A_\perp$ a polarized electron beam of 20~$\mu$A has been
scattered off a 10~cm liquid hydrogen target. The scattered particles
have been detected under a scattering angle of $30^\circ < \theta_e < 40 ^\circ$
in the PbF$_2$-calorimeter, which has a solid angle of \solidangle~sr and measures
the energy of the scattered particles deposited in the 1022 PbF$_2$ crystals.
The detector is $\phi_e$-symmetric around the beam axis. The luminosity
is permanently measured by 8 water-\cherenkov\ detectors located at small
electron scattering angles $4^\circ < \theta_e < 10 ^\circ$,
symmetric around $\phi_e$.
The luminosity monitors have been optimized for the detection of \moeller\ scattering.
The transverse beam spin asymmetry in \moeller\ scattering
is of the same order as in elastic electron proton
scattering \cite{transversemoeller:dixon:2004}. Using the $\phi_e$-symmetry of the luminosity detectors
we average over the 8 luminosity monitors before normalizing  target density fluctuations to
the luminosity signal in the extraction of the asymmetry.\\
We have measured $A_\perp^m$ at two different beam energies of
\beamenergylow~MeV and at \beamenergyhighII~MeV corresponding to an
acceptance averaged four-momentum transfer of \qsquaredaveragedlow~(GeV/c)$^2$
and \qsquaredaveragedhighI~(GeV/c)$^2$, respectively.
The same method of inserting a $\lambda/2$-plate
in the laser system of the source as described in \cite{a4experiment:maas:2004}
has been applied in order to minimize false asymmetries and test for systematic errors.
The transverse beam spin asymmetry and the associated systematic error
has been determined using the
same analysis method after correcting for beam polarization,
target density fluctuations, nonlinearities in the luminosity monitors and
dead time in the calorimeter as in \cite{a4experiment:maas:2004}.
The $\phi_e$ dependence of the measured $A_\perp^m$ leads to a complete
cancelation of the transverse asymmetry if averaged over the $\phi_e$-symmetric
detector. Therefore we have made 8 subsets of the 1022 detector channels of the PbF$_2$-calorimeter,
each subset spanning a sector of 45$^\circ$ in $\phi_e$. The result of our
measurements can be seen in Fig.~\ref{fig:aperp:measurement}.
The data at \beamenergylow~MeV and at \beamenergyhighII~MeV represent
\datahourslow~h and \datahourshigh~h of data taking time, respectively.
One sees a clear $\cos(\phi_e)$-modulation
as expected from $A_\perp^m$ taking into account our
definition of $\phi_e$ in Fig.~\ref{fig:aperp:angle}. The solid lines in
Fig.~\ref{fig:aperp:measurement} represent a fit to the data points
of the form $A_\perp^m = A_\perp \int_{(\phi_e-22.5^\circ)}^{(\phi_e+22.5^\circ)}
\cos(\phi_e') d\phi_e' = 0.765 A_\perp  \cos(\phi_e)$.
Including all corrections, we obtain a value of
$A_\perp$(Q$^2$ = \qsquaredaveragedlow (GeV/c)$^2$)=(
\experimentalasymmetryalulowcorr\ $\pm$ \statisticalerrorlow$_{\rm stat}$ $\pm$ \combinedsyspolerrorlowalucor$_{\rm sys}$)~ppm
and
$A_\perp$(Q$^2$ = \qsquaredaveragedhighII (GeV/c)$^2$)=(
\experimentalasymmetryaluhighcorr\ $\pm$ \statisticalerrorhigh$_{\rm stat}$ $\pm$ \combinedsyspolerrorhighalucor$_{\rm sys}$)~ppm.
The first error represents in both cases the statistical error and the second the systematic uncertainties.
\begin{figure}
  \centering
  \includegraphics[width=0.45\textwidth]{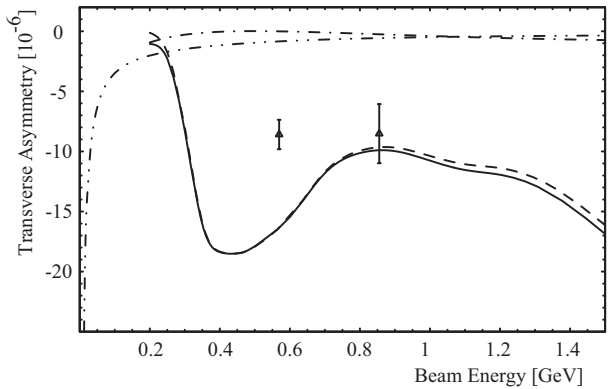}\\
  \caption{The results of two model calculations \cite{2photonamplitude:vanderhaeghen:2003,2photon:musolf:2004}
  are shown with the results of our measurements of $A_\perp$ (see text for explanation).
  }\label{fig:aperp:resultvsenergy}
\end{figure}
In Fig.~\ref{fig:aperp:resultvsenergy} our measured asymmetries
are compared to calculations from
\cite{2photonamplitude:vanderhaeghen:2003}. For the intermediate hadronic state
the ground state proton (elastic contribution, dash-dotted line) has been
used together with excitation amplitudes to $\pi$N-intermediate states
(inelastic contribution, dashed line) as
described by MAID \cite{maid}.
The solid line shows the result
for the full calculation. The dash-double-dotted line represents the results from
a calculation using an effective theory of electrons, protons and photons \cite{2photon:musolf:2004}
which should be compared to the elastic contribution. The data points are the results of our measurement
at \beamenergylow~MeV and at \beamenergyhighII~MeV.
%
%
%
Our measurements of $A_\perp$ clearly show that the two-photon exchange contribution
is already dominated at our low-Q$^2$-kinematics of
Q$^2$=\qsquaredaveragedlow~GeV$^2$ and Q$^2$=\qsquaredaveragedhighII~GeV$^2$ to a large extent by the
inelastic $\pi$N-intermediate state of $\Delta(1232)$-resonance and
higher resonances.\\
The extraction of ${\cal I}(\hat{\mathrm{F}}_{3}(s, Q^2))$,
${\cal I}(\hat{\mathrm{F}}_{4}(s, Q^2))$, and ${\cal I}(\hat{\mathrm{F}}_{5}(s, Q^2))$
from a measurement of $A_\perp$ is in principle possible.
The knowledge of the imaginary part of $\hat{\mathrm{F}}_{3}(s, Q^2)$ can be used
to calculate the real part of $\hat{\mathrm{F}}_{3}(s, Q^2)$, for example
by applying dispersion relations. This would give
the unique possibility of comparing a model calculation
for the real part of $\hat{\mathrm{F}}_{3}(s, Q^2)$ with the extraction
done from the measurement of the imaginary part. Such an experimental verification
of the two-photon contribution to the cross section is at present impossible
due to the lack of data. We plan on a series of measurements of
$A_\perp$ at different beam energies under forward and backward angles \cite{sfbantrag:maas:2004}.
Testing the theoretical framework used to compute them is important for
the interpretation of measurements testing the
Standard Model. In particular this applies to  the neutron $\beta$-decay correlation
experiments. Combined with
the neutron lifetime they allow a determination of the Kobayashi-Maskawa matrix element
$V_{ud}$ that is free from nuclear theory uncertainties.
Similarly, the Q-weak experiment at Jefferson Lab will test the Standard Model running
of $\sin^2\theta_W$. Any
discrepancies between the Standard Model predictions for these quantities
and the experimental values could point to physics beyond the Standard
Model, to the extent that theoretical uncertainties in the Standard Model
radiative corrections can be shown to be sufficiently small.
In addition to the implications
for the electroweak physics and physics beyond the Standard Model,
this opens the possibility to access the doubly virtual Compton
scattering tensor of the neutron by measuring $A_\perp$ on the deuteron.\\
%
%
%
%
%
This work is supported by the Deutsche Forschungsgemeinschaft
in the framework of the SFB 201, SPP 1034, by the IN2P3 of CNRS
and in part by the US Department of Energy.
We are indebted to K.H. Kaiser and the whole MAMI crew for their
tireless effort to provide us with good electron beam. We also
would like to thank the A1 Collaboration for the use of the
\moeller\ polarimeter. We would like to thank M.~Gorshteyn, B.~Pasquini,
M.~Ramsey-Musolf and M.~Vanderhaeghen for useful discussions.

\bibliography{prl_855MeV_trans}

\end{document}